# Quantum probe for molecular synthesis at room temperature

Susan Z. Hua

*Bio-MEMS and Bio-Materials Laboratory, Materials Program, Mechanical & Aerospace Engineering Department, SUNY-Buffalo, Buffalo, NY 14260*

Harsh Deep Chopra

*Thin Films and Nanosynthesis Laboratory, Materials Program, Mechanical & Aerospace Engineering Department, SUNY-Buffalo, Buffalo, NY 14260*

**Abstract**

In the present study, a prototypical reaction in biochemistry involving a well-known bio-molecule called imidazole ($C_3H_4N_2$) and its known affinity towards transition metals (such as Co, Ni, Cu, Fe, etc.) has been used to illustrate and exemplify the use of ballistic conductors as probes for molecular reactions at the quantum level. Atomic point contacts of transition metal Co were made using electro-deposition, followed by their reaction with imidazole in an aqueous solution. As the imidazole molecules covalently bond with the Co atoms, the quantized conductance of the nanocontact decreases, thus providing a quantum probe for studying molecular reactions. During the course of the reaction, different conductance channels within the nanocontact become unavailable due to covalent bonding of Co with imidazole, thereby also providing a means of studying the number of channels contributing to the total conductance in the nanocontact itself.

Classical or diffusive electron transport is characteristics of metallic conductors having physical dimensions that are large in comparison to the mean free path of the electrons. As the size of the conductor becomes comparable to or smaller than the mean free path, the transport enters the ballistic regime. In contrasts to the zigzag motion of the electrons in a diffusive conductor, ballistic transport is characterized by bullet-like motion through the conductor devoid of any scattering events. Sharvin has derived the resistance of a ballistic conductor using the semi-classical approach;[1] see also Ref. [2] for an equivalent treatment by Wexler who used a model analogous to effusion of gases across a pinhole first studied by Knudsen.[3] In metals, when the ballistic conductor is made of a single atom or at most a few atoms, its size becomes comparable to the Fermi wavelength (typically ≤0.5 nm in metals). Such an atomic point contact may be viewed as a 'waveguide' for the electrons and the classical picture of electron transport is replaced by Landauer's seminal concept of conduction as a transmission probability for the propagating electron waves across the waveguide.[4] The transverse confinement of the electron wave functions quantizes the energy levels within such a nanocontact to form bands or 'channels', with each channel contributing to conductance in discrete units of $2e^2/h$ ($\equiv 1G_o$); $e$ is the charge of the electron and $h$ is the Planck's constant. The total conductance is obtained by summing over all the open channels, and is given by the well-known Landauer-Büttiker equation, $G = (2e^2/h)\sum_{n=1}^{N} T_n$.[4,5] Here $T_n$ is the transmission probability of the $n^{th}$ channel (assuming spin degeneracy for the electrons). The value of $T_n$ can range from 0 to 1, corresponding to a fully closed or a fully opened channel, respectively. Values of $T_n$ intermediate between 0 and 1 reflect a partially open channel, which among other factors also depend on the geometry of the nanocontact.[6] Ideally, when the channels are fully open, the conductance changes in a stepwise fashion during the growth of the nanocontact, with step height equal to $1G_o$. However, in case of ferromagnetic materials, the spin degeneracy can be lifted (by making the nanocontacts in the presence of an applied magnetic field), in which case the stepwise change in conductance occurs in units of $\frac{1}{2}G_o$ instead of $1G_o$.[7-8]

Previously, the above described principles of ballistic conductance have been used for the detection of different molecules in a solution (a chemical sensor) and towards the goal of making molecular electronics.[9-10] In the present study, ballistic conductors are used as probes for molecular synthesis at the quantum level, whereby the change in conductance due to the formation of a transition metal-complex on the surface of a ballistic conductor is studied. Specifically, a widespread reaction in biochemistry involves the reaction of a bio-molecule called imidazole ($C_3H_4N_2$) with a transition metal (TM) such as Co, Ni, Fe, Cu, etc. to form transition a metal complex $TM(C_3H_4N_2)$. Imidazole is an important heterocyclic bio-molecule and is also part of an amino acid called Histadine.[11] Imidazole often plays a key role in site activity of proteins and enzymes. Due to its affinity to form transition metals complex, it is also widely used to displace peptides bound to a column of transition metal ligands.[12-13] In the present study, the stepwise (or quantized) change in conductance of a Co nanoconductor was studied during the course of its reaction with imidazole molecules to form the metal complex $Co(C_3H_4N_2)$; other transition metals react similarly with imidazole.

The Co nanocontacts were electro-deposited between microfabricated Co electrodes made by standard photolithographic processing on oxidized silicon wafers. As shown in Fig. 1(a), the electrodes had a gap of 50-300 nm, and were 2-3 µm thick. This was followed by room temperature electro-deposition of Co between the electrodes using a cobalt sulfamate solution (solution pH was 3.3). The deposition voltage was typically between 0.7 to 1 V.

Control Over Contact Size: Recently Boussaad and Tao have described a simple and elegant self-terminating electro-deposition technique to make nanocontact and nano-gaps of controlled dimensions.[16] Their technique allows stable atomic contacts to be made repeatedly. This method has been previously used by the authors to make Ni and Co nanocontacts,[8,17] and was used to make the nanocontacts in the present study. The experimental layout of the self-terminating electrochemical method is shown in Fig. 1(b), which shows an external resistor $R_{ext}$ connected in series with the electrochemical cell; $R_{ext}$ is set to a value that is less than the quantum resistance $1/G_o$ of 12900 Ω, and was typically 6500 to 12900 Ω in the present study. In Fig. 1(b), let the resistance between the anode and the cathode be $R_{cell}$, (the typical cell resistance was several MΩ). Therefore the voltage across the gap between the anode and the cathode is given by $V_{cell} = [R_{cell}/(R_{cell} + R_{ext})]V_o$, where $V_o$ is the applied voltage. Initially, when the gap between the electrodes is large $R_{cell} >> R_{ext}$ and $V_{cell} \approx V_o$. As soon as the contact forms between the two electrodes, the applied voltage across the electrodes drops and gets established across the external resistor $R_{ext}$. As a result the growth of the nanocontact slows considerably at which point the electro-deposition is stopped.

Figure 2 shows the conductance plots for five different nanocontacts, stabilized at values ranging from $1G_o$ to $5G_o$, made by using the above-described self-terminating method. In Fig. 2, the stabilization of nanocontacts at different conductance plateaus reflects a change in the number of channels contributing to the total conductance, which, in turn, is a function of the contact diameter. As the nanocontact diameter decreases, the separation between different energy bands (or channels) increases due to an increase in the transverse confinement of the wave functions of the propagating electron waves. This causes some of the channels to rise above the Fermi level, thereby becoming unavailable for conductance. In the limit of a single atom contact, only one channel may lie below the Fermi level, for a total conductance of just $1G_o$. Figure 2 also shows that the Co nanocontacts not only exhibit integral values of $G_o$ but also non-integral values of $1.3G_o$ and $2.6G_o$. A non-integral value of conductance reflects transmission probability $T_n$ (in the Landauer-Büttiker equation $G = G_o \sum_{n=1}^{N} T_n$) for each channel intermediate between 0 and 1, and this will be discussed later. The inset in Fig, 2 shows the conductance plot for a Co nanocontact deposited in the presence of an applied magnetic field (1800 Oe). Growth of a nanocontact in the presence of an applied magnetic field lifts the spin degeneracy of the ferromagnetic Co, resulting in step-wise change in conductance in units of $\frac{1}{2}G_o$ instead of $1G_o$, as shown in the inset in Fig. 2 ($T_n$ equal to 1 in the Landauer-Büttiker equation assumes spin degeneracy, absence of which causes spin splitting). This behavior is similar to the previously observed behavior of Ni nanocontacts made in the presence of an applied magnetic field.[7,8]

Following the formation of a stable Co nanocontact, the electrolyte was gradually replaced by de-ionized ultra-filtered (D.I.U.F.) water by repeated purging of the electrolytic cell with D.I.U.F. water. This was followed by the dissolution of a small amount of imidazole crystals (in μM/liter quantities) in water. The dissolved imidazole gradually reacts with Co atoms within the nanocontact as well as the Co atoms on the surface of the electrodes to form the metal complex $Co(C_3H_4N_2)$. The corresponding change in the nanocontact conductance as a function of time is shown in Fig. 3. The Co nanocontact in Fig. 3 had a stable conductance of $2.6G_o$ and was highly stable for over an hour prior to the start of the chemical reaction (a portion of the recorded conductance plot for this sample is shown in Fig. 2). Figure 3 shows that following the dissolution of the imidazole molecule in water, the conductance decreases in four stages, marked $G_1$ through $G_4$. Remarkably, each stage is characterized by an equal decrease in conductance of roughly $\frac{2}{3}G_o$. During the first and the third stage ($G_1$ and $G_3$), the decrease in conductance is gradual, while the second and the fourth stage ($G_2$ and $G_4$) is marked by an abrupt decrease in conductance. The gradual decrease in the conductance (stages 1 and 3) reflects covalent bonding of the imidazole molecules with the Co atoms in the vicinity of the nanocontact, as shown schematically in Fig. 3. Covalent bonding of the Co atoms with the imidazole molecules (imidazole molecules indicated by 'Y' in the schematic in Fig. 3) effectively reduces the cone angle at the nanocontact, and thus the transmission probability,[6] as shown in the schematic in Fig. 3. The gradual change in conductance in stages 1 and 3 in Fig. 3 is thus indicative of contribution to the total conductance from Co atoms that are adjacent to the nanocontact. Eventually, by probability, an imidazole molecule reacts with Co atom within the nanocontact. This localization of the electrons in the nanocontact due to covalent bonding with the imidazole molecules is reflected by an abrupt lifting of the energy band above the Fermi level and a corresponding decrease in the total conductance. Ultimately, when the reactant imidazole molecule blocks all the channels within the nanocontact, the conductance drops to zero.

Finally, the occurrence of non-integral contribution to the total conductance from different channels has been previously shown (both experimentally and theoretically) in complex metals involving $p$ and $d$ electrons (where the hybridization of the $s$ electrons with $p$ or $d$ electrons results in the formation of a large number of channels).[18-19] For example, hybridization of the $s$ and $p$ electrons in aluminum results in three channels contributing $\frac{1}{3}G_o$ per channel.[16] Present results show that in case of Co (where $s$ and $d$ electrons are present), more than one channels may contribute to the total conductance as seen from multiple steps in Fig. 3. Therefore, the present method also provides a means of studying the channels contributing to the overall conductance in a given metal system.

This work was supported by the Nationals Science Foundation, grants NSF-DMR-, NSF-DMR-97-31733. S.Z.H. also acknowledges the support of NSF-CMS-02-01293. The authors also acknowledge Fred Sachs from the Department of Physiology and Biophysics at SUNY-Buffalo for useful discussions, Matthew Sullivan and Dan Ateya for preparation of the microfabricated templates for the samples at the Cornell Nanofabrication Facility (a member of the National Nanofabrication Users Network), which is supported by NSF Grant ECS-9731293, Cornell University and industry affiliates.

# FIGURE CAPTIONS

**FIG. 1.** (a) Scanning electron microscope of a microfabricated template between which Co nanocontacts were electro-deposited. (b) Schematic of the self-terminating method to make atomic sized nanocontacts.

**FIG. 2.** Conductance plots versus time for different Co nanocontacts stabilized at different values of $G_o$. The inset shows a conductance plot for a Co nanocontact deposited in 1800 Oe constant field, which causes spin splitting of the electrons and stepwise change in conductance in units of $\frac{1}{2}G_o$ instead of $G_o$.

**FIG. 3.** Conductance plot of a Co nanocontact during its reaction with imidazole to form the metal complex.

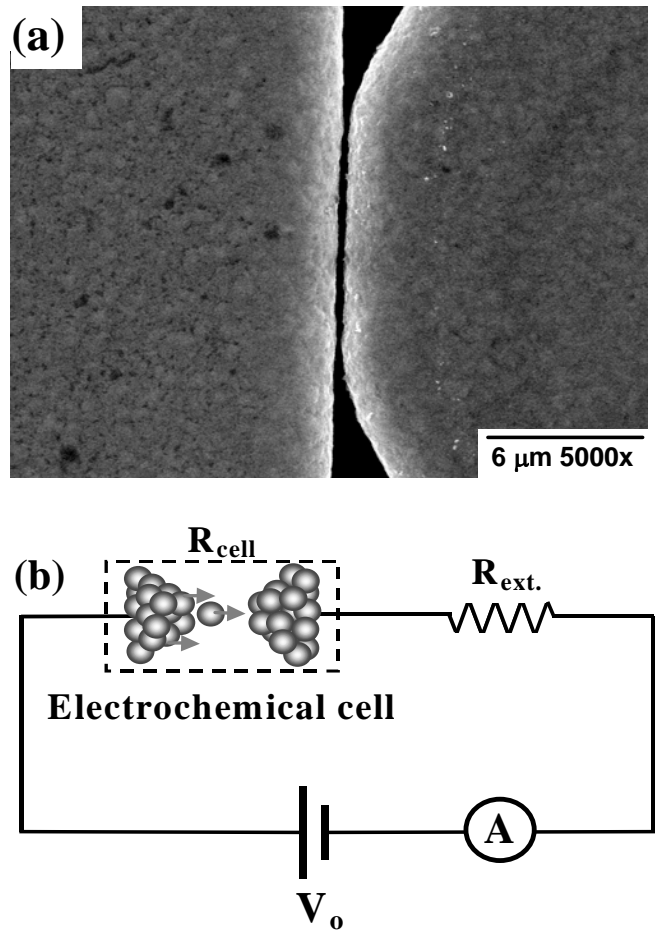

**Figure 1**
Hua and Chopra (Phys. Rev. B – Rapids)

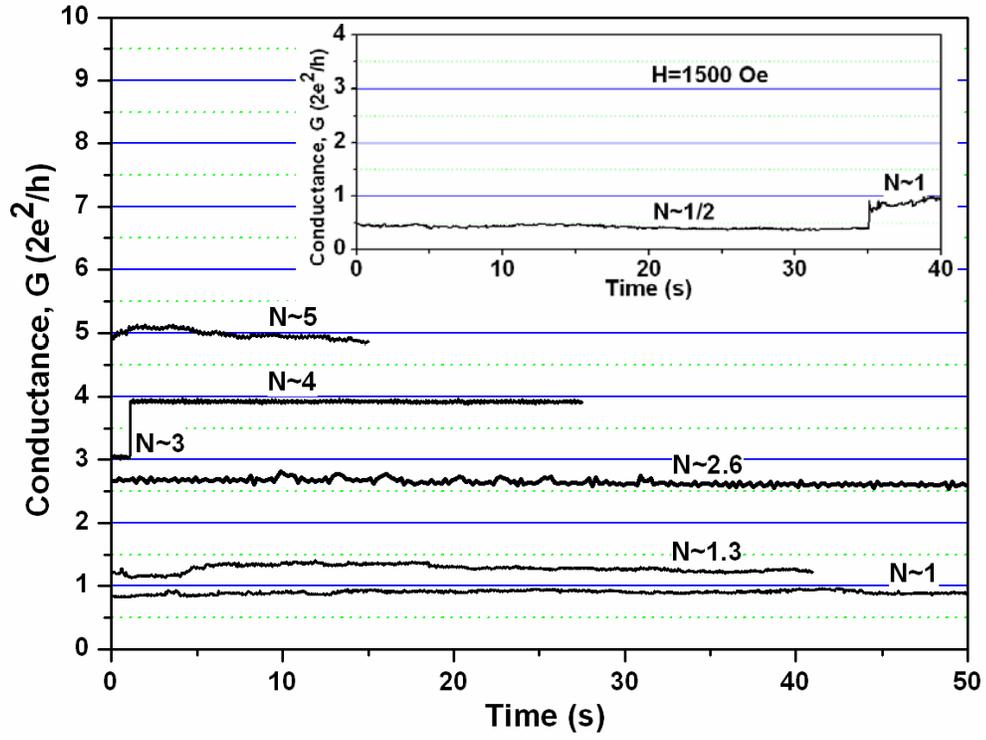

**Figure 2**
Hua and Chopra (Phys. Rev. B – Rapids)

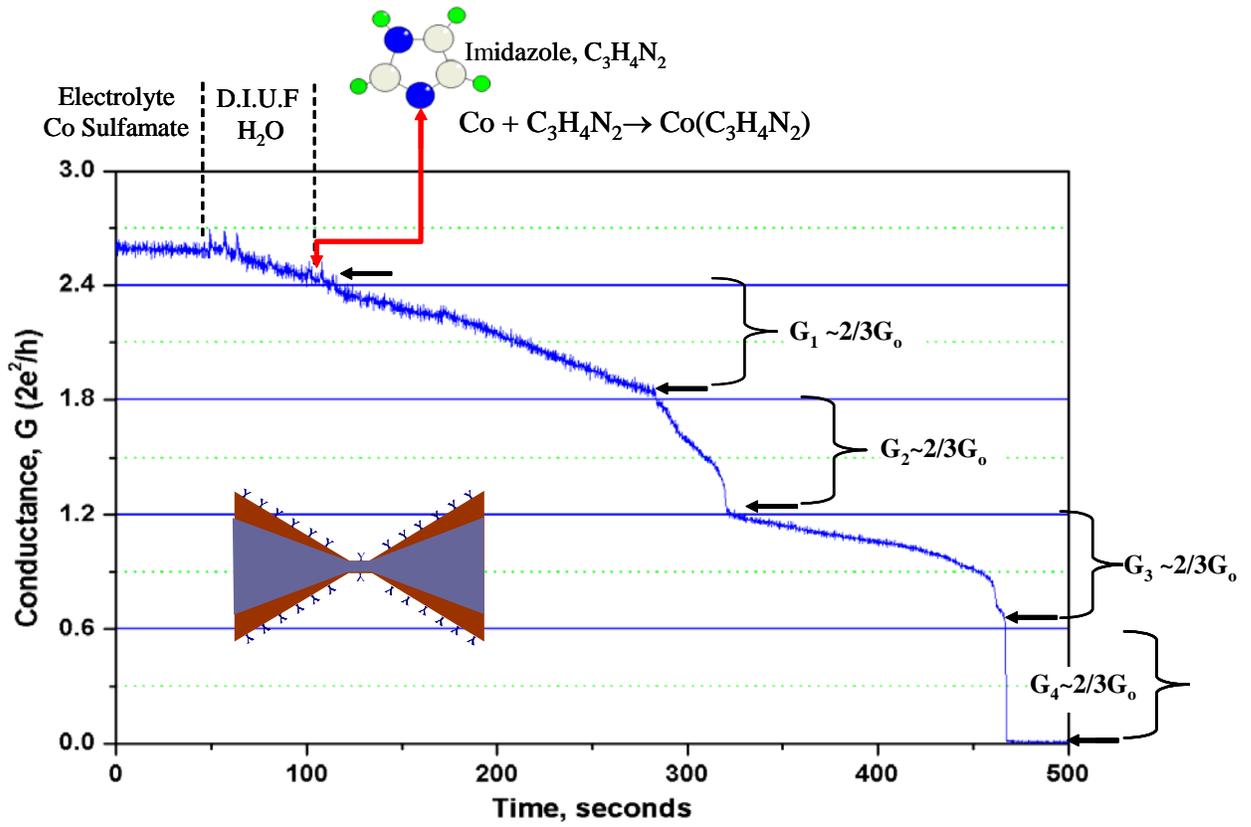

**Figure 3**
Hua and Chopra (Phys. Rev. B – Rapids)